\documentclass[aps,prd,twocolumn,showpacs,preprintnumbers,nofootinbib,amsmath,amssymb]{revtex4}

\usepackage{graphicx}
\usepackage{subfigure}
\usepackage{epstopdf}
\usepackage{dcolumn}
\usepackage{amsmath}
\usepackage[normalem]{ulem}

\usepackage{blindtext}
\usepackage{xcolor}

\begin{document}

\title{Screening masses in strong external magnetic fields}

\author{Claudio Bonati}\email{claudio.bonati@unifi.it}
\affiliation{Dipartimento di Fisica e Astronomia dell'Universit\`a di
  Firenze and INFN - Sezione di Firenze,\\ Via Sansone 1, 50019, Sesto
  Fiorentino (FI), Italy.}

\author{Massimo D'Elia}\email{massimo.delia@unipi.it}
\author{Marco Mariti}\email{mariti@df.unipi.it}
\author{Michele Mesiti}\email{michele.mesiti@pi.infn.it}
\author{Francesco Negro}\email{fnegro@pi.infn.it}
\author{Andrea Rucci}\email{andrea.rucci@pi.infn.it}
\affiliation{Dipartimento di Fisica dell'Universit\`a di Pisa and INFN
  - Sezione di Pisa,\\ Largo Pontecorvo 3, I-56127 Pisa, Italy.}

\author{Francesco Sanfilippo}\email{francesco.sanfilippo@roma3.infn.it}
\affiliation{INFN, Sezione di Roma Tre, Via della Vasca Navale 84, I-00146 Rome, Italy.\\}

\date{\today}

\begin{abstract}
We present results for the (color)magnetic and (color)electric screening masses
of the Quark-Gluon Plasma in the presence of an external magnetic field.  The
screening
masses are extracted from the correlators of Polyakov loops, determined
by lattice QCD 
simulations at the physical point. We explore temperatures in the range
$200\,\mathrm{MeV}\lesssim T\lesssim 330\,\mathrm{MeV}$ and
magnetic field
intensities up to $|e|B \sim 1.3\,\mathrm{GeV}^2$.  We find that both
screening masses are increasing functions of the magnetic field and that the
dependence on $B$ becomes weaker for larger temperatures. In the
case of the magnetic screening mass a slight anisotropy is also observable.
\end{abstract}


\maketitle

\section{Introduction}
\label{intro}

The fate of heavy quark bound states as a probe of the deconfining properties
of the thermal strongly interacting medium has a long history. In the seminal
paper Ref.~\cite{matsui} a suppression of the production rate of these states
was predicted, being caused by the shortening of the screening length for color
interactions in the Quark-Gluon Plasma.  In subsequent analyses it was realized
that the situation is in fact more involved: various effects can enhance or
suppress this phenomenon, the final result being the outcome of a complex
dynamical process (see Ref.~\cite{andronic} for a recent review of the
theoretical and experimental aspects). Here we consider the consequences that
the introduction of a magnetic background field may have on the screening
lengths of the thermal medium.

The presence of strong magnetic backgrounds, with field strengths comparable to
the QCD scale, is a situation common to many contexts, ranging from
cosmology~\cite{vacha,grarub} and non-central heavy ion collisions~\cite{hi1,
hi2, hi3, hi4, tuchin,Holliday:2016lbx}, with magnetic fields going up to
$10^{16}$ Tesla ($e B \sim 1$~GeV$^2$), to magnetars~\cite{magnetars}, where
magnetic fields of the order of $10^{11}$ Tesla are expected on the surface,
and possibly higher in the inner core.

How the properties of strongly interacting matter are modified by such magnetic
fields has been the subject of many theoretical efforts, see
Refs.~\cite{Kharzeev:2012ph, Miransky:2015ava} for recent reviews. As regards 
the
effects more directly related to color interactions, various studies have
considered the possible influence of an external magnetic field on the static
quark-antiquark potential~\cite{Miransky:2002rp, Chernodub:2014uua,
Rougemont:2014efa, Ferrer:2014qka, Simonov:2015yka, DElia:2015eey}, which has been clarified
by recent lattice results~\cite{nostro1,nostro2}, and might have consequences
relevant to the spectrum of heavy quark bound states~\cite{Alford:2013jva,
Dudal:2014jfa, Cho:2014loa, Taya:2014nha, Bonati:2015dka, Hattori:2015aki,
Suzuki:2016kcs, Guo:2015nsa, Fukushima:2015wck, Gubler:2015qok,Hasan:2017fmf}. At zero
temperature, the potential becomes anisotropic and the string tension $\sigma$
is larger (smaller) in the direction orthogonal (parallel) to the magnetic
field $\mathbf{B}$~\cite{nostro1,nostro2}; at finite $T$, in particular in the
region right below the pseudocritical temperature $T_c$, the magnetic field
induces a general suppression of $\sigma$~\cite{nostro2}, leading to an early
onset of deconfinement, in agreement with the observed dependence of $T_c$ on
$B$~\cite{Bali:2011qj, Bruckmann:2013oba, Ilgenfritz:2013ara}

In this paper we extend the study to the region of temperatures above $T_c$, in
order to investigate the effects of a magnetic background on the interactions
between heavy quarks in the Quark-Gluon Plasma.  In this phase, the effective
interaction is no longer confining and can instead be described by a screened
Coulomb form, with two different screening lengths/masses characterizing the
(color)electric and the (color)magnetic sectors.  Studying the appropriate
combinations of Polyakov loop correlators by means of Lattice QCD simulations
at the physical point we estimated the screening lengths for $T\simeq 200, 250,
330\,\mathrm{MeV}$ and several values of the external magnetic field (for
$|e|B\lesssim 1.3\,\mathrm{GeV}^2$).  From these results we conclude that the
magnetic field induces a reduction of both screening lengths (i.e. an increasing
of the screening masses), which might induce a further suppression of the
formation of heavy quark bound states in the thermal medium.

The paper is organized as follows. In Section II we discuss our numerical setup
and review the definition of the screening masses in terms of Polyakov loop
correlators.  In Section III we present our numerical results. Finally, in
Section IV, we draw our conclusions.

\section{Setup}

\subsection{Physical observables}
\label{physobs}

The screening masses of strongly interacting matter have been historically
introduced in perturbation theory by studying the pole structure of the finite
temperature gluon propagator. While this approach presents no difficulties for
the computation of the leading order electric screening mass (see e.g.
\cite{Gross:1980br, KG_book, Laine_book}), it was soon realized that perturbation theory
gets into trouble at higher orders, or even at the 
leading order for the magnetic
mass, because of the infrared sensitivity of the obtained expressions
\cite{Nadkarni:1986cz, Rebhan:1993az, Rebhan:1994mx, Braaten:1994pk, KG_book, Laine_book}.

The natural way to overcome these difficulties and obtain nonperturbative
results for the screening masses is to analyze the large distance behavior of
gauge-invariant correlation functions.  For this purpose, correlators of
Polyakov loops have been traditionally used \cite{Braaten:1994qx,
Arnold:1995bh} and the two independent correlators that can be studied are
\begin{equation}\label{eq:polycorr}
  \begin{aligned}
  & C_{LL^\dag}(\mathbf{r},T) = \Big\langle\mathrm{Tr}L(\mathbf{0})\,\,\mathrm{Tr}L^\dag(\mathbf{r})\Big\rangle\\
  & C_{LL\ }(\mathbf{r},T) = \Big\langle\mathrm{Tr}L(\mathbf{0})\,\, \mathrm{Tr}L (\mathbf{r})\Big\rangle\ ,
  \end{aligned}
\end{equation}
where $L(\mathbf{x})$ is the Polyakov loop operator, which is defined in
the continuum by
\begin{equation}\label{eq:polyloop}
  L(\mathbf{x})=\frac{1}{N_c}\mathcal{P}\exp\left(-ig\int_0^{1/T} 
    A_0(\mathbf{x},\tau)\mathrm{d}\tau\right)\ .
\end{equation}
In this expression $N_c$ is the number of colors and the symbol
$\mathcal{P}\exp$ denotes the path-ordered exponential.

The correlator $C_{LL^\dag}$ is often studied because of its connection
with the free energy $F_{Q\bar{Q}}(\mathbf{r},T)$ of a static quark-antiquark
pair, that can be computed using the relation \cite{McLerran:1981pb} 
\begin{equation}
F_{Q\bar{Q}}(\mathbf{r},T)=-T\log C_{LL^\dag}(\mathbf{r},T) \, .
\end{equation}
The study of $C_{LL^{\dag}}$ is thus the finite temperature
counterpart of the study of Wilson loops at zero temperature, from
which the potential energy of a static quark-antiquark pair can be
extracted. Since in any numerical study the temporal extent of the
lattice is always finite, it would also be possible to extract the
static potential as the zero temperature limit of the free energy
extracted from Polyakov loops, however this second procedure is
generically not numerically convenient because of the much larger
statistical errors involved.

\begin{figure}[t]
  \includegraphics[width=\columnwidth,clip]{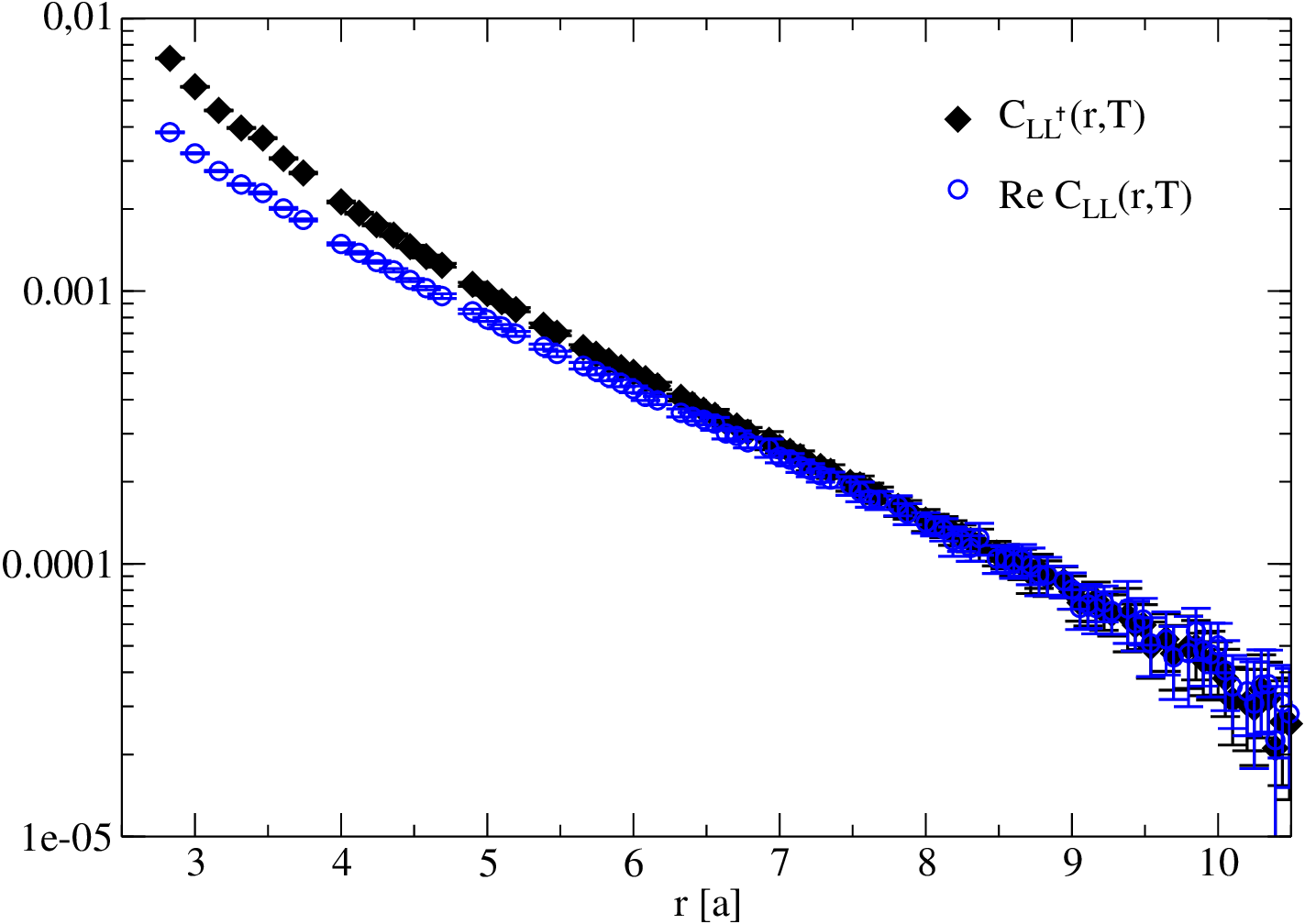}
  \caption{Comparison of the correlator $C_{LL^{\dag}}(r,T)$ and the
    real part of the correlator $C_{LL}(r,T)$ for
    $T\simeq 200\,\mathrm{MeV}$ (in both cases we considered the
    connected correlators). Only the case $B=0$ is reported for
    clarity but also in the case of non-vanishing external magnetic
    field the behaviour of the correlators is qualitatively similar.}
  \label{fig:corrsLLandLLdag}
\end{figure}

In the following we will need also the somehow less used correlator
$C_{LL}(\mathbf{r},T)$; a comparison of the behaviour of
$C_{LL^{\dag}}(r,T)$ and $\mathrm{Re}\,C_{LL}(r,T)$ for
$T\simeq 200\,\mathrm{MeV}$ is shown in
Fig.~\ref{fig:corrsLLandLLdag}, from which it is clear that the very
large distance behavior of these two correlators is the same. In fact,
the two correlators turn out to be substantially different only in the
confined phase, where the $C_{LL}(\mathbf{r},T)$ correlator is
strongly suppressed also at short distances because of the confining
properties of the medium\footnote{Actually, in the pure gauge theory,
  $C_{LL}(\mathbf{r},T)$ vanishes exactly at all distances in the
  confined phase, because of the exact center symmetry.}.

Under Euclidean-time reversal $\mathcal{R}:\tau\to-\tau$ the color-magnetic and
color-electric gluon components, $A_i(\mathbf{x},\tau)$ and
$A_4(\mathbf{x},\tau)$, are respectively even and odd.  Since the Polyakov loop
transforms under such a transformation as $\mathcal{R}:L\to L^\dag$, the
following combinations
\begin{equation}
  L_{M}=\frac{1}{2}\left(L+L^\dag\right) \qquad L_{E}=\frac{1}{2}\left(L-L^\dag\right)
\end{equation}
receive contributions only from the magnetic and electric sectors respectively.
These quantities can be further decomposed in eigenstates of the charge
conjugation operator $\mathcal{C}$: under this symmetry $\mathcal{C}:L\to L^*$
and we thus obtain
\begin{equation}\label{LEMpm}
  L_{M^\pm}=\frac{1}{2}\left(L_M\pm L_M^*\right) \qquad 
  L_{E^\pm}=\frac{1}{2}\left(L_E\pm L_E^*\right) \,
\end{equation}
where the subscripts $\pm$ indicates the $\mathcal{C}$ eigenvalues.  Such a
decomposition has been introduced in Ref.~\cite{Arnold:1995bh} and it has been
recently used in some lattice computations (see e.g.
\cite{Maezawa:2010vj,Borsanyi:2015yka}).

From the definitions in Eq.~\eqref{LEMpm} it immediately follows that
the magnetic odd and the electric even sectors are trivial, i.e.
\begin{equation}
\mathrm{Tr}L_{M^-}=\mathrm{Tr}L_{E^+}=0\ ,
\end{equation}
and that the following relations hold true
\begin{align}
& \mathrm{Tr}L_{M^+}=\mathrm{ReTr}L \label{prop1} \\
& \mathrm{Tr}L_{E^-}=i\mathrm{ImTr}L \label{prop2}\ .
\end{align}
Using the previous relations, the correlators needed to investigate the
magnetic and electric sectors can be written in the form
\begin{equation}\label{eq:mecorrs}
  \begin{aligned}
    C_{M^+}(\mathbf{r},T) & 
    =\Big\langle\mathrm{Tr}L_{M^+}(\mathbf{0})\mathrm{Tr}L_{M^+}(\mathbf{r})\Big\rangle
    -\big|\big\langle\mathrm{Tr}L\big\rangle\big|^2\\
    C_{E^-}(\mathbf{r},T) &
    =-\Big\langle\mathrm{Tr}L_{E^-}(\mathbf{0})\mathrm{Tr}L_{E^-}(\mathbf{r})\Big\rangle
  \end{aligned}
\end{equation}
where the minus sign in the definition of $C_{E^-}$ is conventional and in the
electric case no disconnected term is present because of the symmetry under
charge conjugation. It is convenient to rewrite the previous correlators in term 
of $C_{LL}$ and $C_{LL^{\dag}}$ and the final result is
\begin{equation}\label{eq:polycorrequivalent}
  \begin{aligned}
    C_{M^+}=&+\frac{1}{2}\mathrm{Re}\big[C_{LL}+C_{LL^\dag}\big]-
\left|\langle \mathrm{Tr}L\rangle \right|^2\\
    C_{E^-}=&-\frac{1}{2}\mathrm{Re}\big[C_{LL}-C_{LL^\dag}\big] \; .
\end{aligned}
\end{equation}
Notice that the mixed electric-magnetic correlator
\begin{equation}
\Big\langle\mathrm{Tr}L_{M^+}(\mathbf{0})\mathrm{Tr}L_{E^-}(\mathbf{r})\Big\rangle \propto 
\Big\langle\mathrm{ReTr}L(\mathbf{0})\, \mathrm{Im Tr}L(\mathbf{r})\Big\rangle
\end{equation}
is zero by charge conjugation invariance, meaning that the real and imaginary 
parts of the Polyakov loop fluctuate independently.

The electric and magnetic screening masses can now be extracted from the large
distance behaviour of the correlators in Eqs.~\eqref{eq:mecorrs}. In
particular, in the very high temperature regime, one expects
\cite{Braaten:1994qx, Arnold:1995bh} these quantities to scale as
\begin{equation}\label{eq:mecorrsfit}
  \begin{aligned}
    C_{E^-}(\mathbf{r},T)\big|_{r\to\infty}\simeq\frac{e^{-m_E(T)r}}{r}\\
    C_{M^+}(\mathbf{r},T)\big|_{r\to\infty}\simeq\frac{e^{-m_M(T)r}}{r}
  \end{aligned}
\end{equation}
where $r=|\mathbf{r}|$ and $m_E(T)$ and $m_M(T)$ are the electric and magnetic
screening masses respectively.  In the subsequent analysis, we will use these
expressions to extract the screening masses from the correlators computed on
the lattice.  In the presence of an external magnetic field, we obviously have to
take into account the explicit breaking of the rotational symmetry: correlators
taken along directions parallel or orthogonal to the magnetic field cannot be
averaged, and two \emph{a priori} different electric screening masses and
magnetic screening masses have to be defined.

\subsection{Numerical setup}
\label{numsetup}

In this work we have adopted a discretization of $N_f=2+1$ QCD based on the
Symanzik tree-level improved gauge action and the stout smeared rooted
staggered action for the fermionic sector. The partition function in the
presence of a magnetic background $B$ is written as
\begin{equation}\label{partfunc}
Z(B) = \int \!\mathcal{D}U \,e^{-\mathcal{S}_{Y\!M}}
\!\!\!\!\prod_{f=u,\,d,\,s} \!\!\!
\det{({D^{f}_{\textnormal{st}}[B]})^{1/4}}\ ,
\end{equation}
where
$\mathcal{D}U$ is the functional integration over the $SU(3)$
gauge link variables, $\mathcal{S}_{Y\!M}$ stands for the tree-level
improved action~\cite{weisz, curci}:
\begin{equation}\label{tlsyact}
\mathcal{S}_{Y\!M}= - \frac{\beta}{3}\sum_{i, \mu \neq \nu} \left(
\frac{5}{6} W^{1\!\times \! 1}_{i;\,\mu\nu} - \frac{1}{12}
W^{1\!\times \! 2}_{i;\,\mu\nu} \right),
\end{equation}
where $W^{1\!\times \! 1}_{i;\,\mu\nu}$ and $W^{1\!\times \!  2}_{i;\,\mu\nu}$
denote, respectively, the real part of the trace of $1\!\times \! 1$ and
$1\!\times \!2$ loops.  Finally, the staggered fermion matrix
\begin{equation}\label{rmmatrix}
\begin{aligned}
(D^f_{\textnormal{st}})_{i,\,j} =\ & am_f
  \delta_{i,\,j}+\!\!\sum_{\nu=1}^{4}\frac{\eta_{i;\,\nu}}{2}
  \left(u^f_{i;\,\nu}U^{(2)}_{i;\,\nu}\delta_{i,j-\hat{\nu}}
  \right. \nonumber\\ &-\left. u^{f*}_{i-\hat\nu;\,\nu}U^{(2)\dagger}_{i-\hat\nu;\,\nu}\delta_{i,j+\hat\nu}
  \right)
\end{aligned}
\end{equation}
is written in terms of two times stout-smeared gauge links
$U^{(2)}_{i;\,\mu}$~\cite{morning}, with an isotropic smearing
parameter $\rho=0.15$, and the $U(1)$ parallel transporters
$u^f_{i;\,\mu}$, which takes the presence of the external
electromagnetic field into account; in both cases the latin indices
denote the position in the lattice and the greek indices denote the
direction of the link.

For a constant and uniform magnetic background directed along the
$\hat{z}$ direction, a possible choice of the $U(1)$ phases is ($q_f$
is the fermion charge)
\begin{eqnarray}\label{bfield}
u^f_{i;\,y}=e^{i a^2 q_f B_{z} i_x} \ , \quad
{u^f_{i;\,x}|}_{i_x=L_x}=e^{-ia^2 q_f L_x B_z i_y}\, ,
\end{eqnarray}
while all the other $U(1)$ link variables are set  to 1.  For this choice to
actually describe a uniform magnetic field on the lattice with periodic
boundary conditions, it is necessary for the value of $B_z$ to satisfy the
following quantization condition \cite{thooft, bound3, wiese}
\begin{equation}\label{bquant}
eB_z={6 \pi b_z}/{(a^2 N_x N_y)} \, ; \ \ \ \ \  b_z \in \mathbb{Z} \, .
\end{equation}

In our numerical simulations, we used for the bare parameters the
values $\beta=3.85$, $m_s/m_l=28.15$ and $am_s=0.0394$, which
correspond \cite{physline1, physline2, physline3} to a lattice spacing
$a\simeq0.0989~\rm{fm}$ and to physical values of the pion and strange
quark masses (isospin breaking is neglected). We performed simulations
on $48^3 \times N_t$ lattices, with $N_t=6,8,10$, corresponding to a
fixed spatial size of around 5 fm and to temperatures
$T\simeq 330~\rm{MeV}, 250~\rm{MeV}, 200~\rm{MeV}$. Polyakov loop
correlators have been measured on a set of around $5\times 10^3$
configurations for each temperature (with measures separated by 5
molecular dynamics trajectories) and, to reduce the statistical noise,
a single step of HYP smearing \cite{Hasenfratz:2001hp} has been
applied to the temporal links, with the parameters of the HYP2-action,
see Ref.~\cite{Della Morte:2005yc}.  Correlators have been extracted
for generic orientations (i.e. not just along the lattice axes) at
$B = 0$, while in the presence of the background field we have
considered separately correlators along the $z$ axis (i.e. parallel to
$B$) and in the whole $xy$ plane (i.e. orthogonal to $B$), which in
the following will be denoted respectively by $Z$ and $XY$.  Note
that, since Polyakov loops renormalize multiplicatively and no further
distance-dependent renormalization enters the correlator, the
screening masses defined by Eq.~\eqref{eq:mecorrsfit} do not need any
renormalization.

\section{Results}\label{res}

\begin{figure}[t]
  \subfigure{\includegraphics[width=\columnwidth,clip]{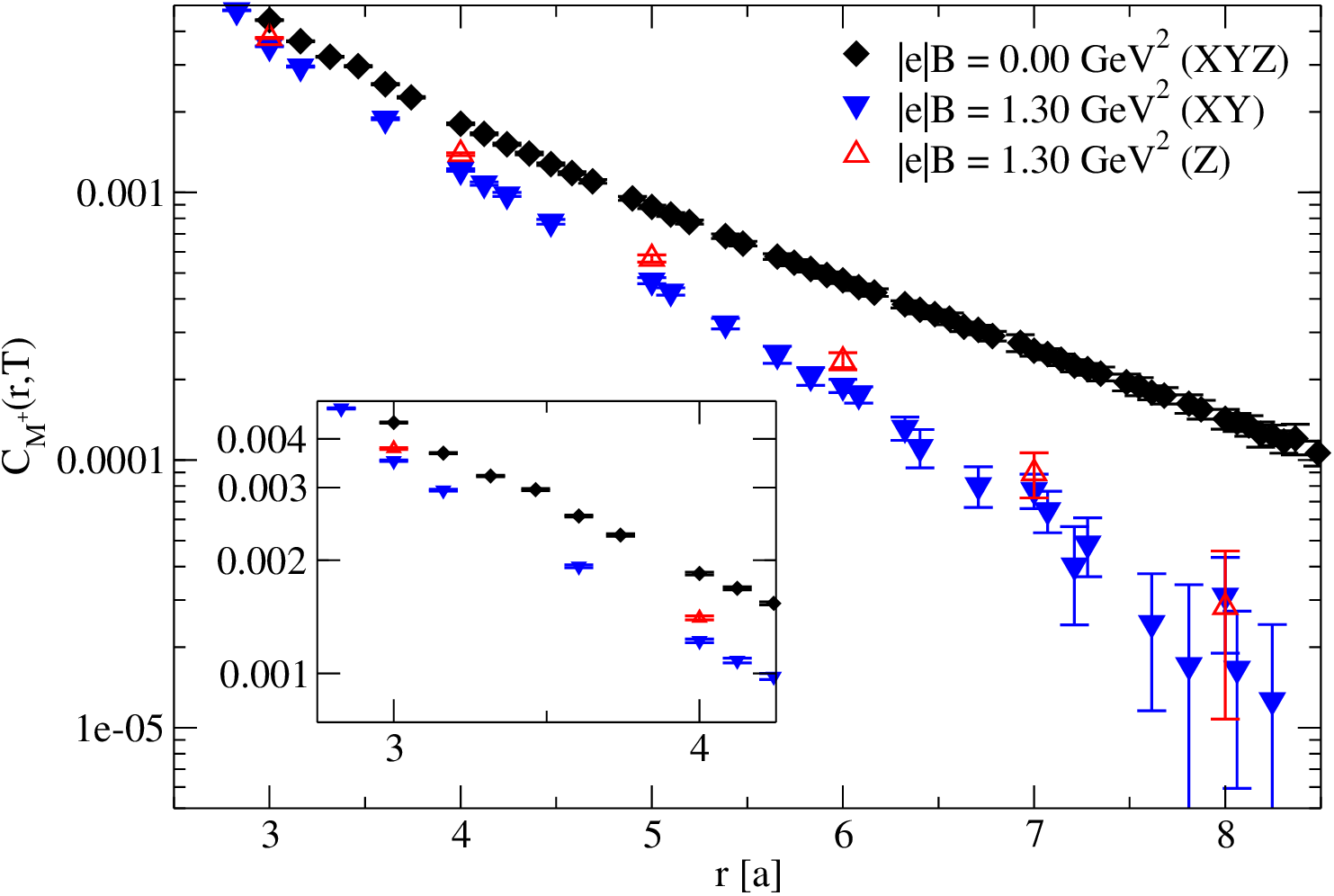}}
  \subfigure{\includegraphics[width=\columnwidth,clip]{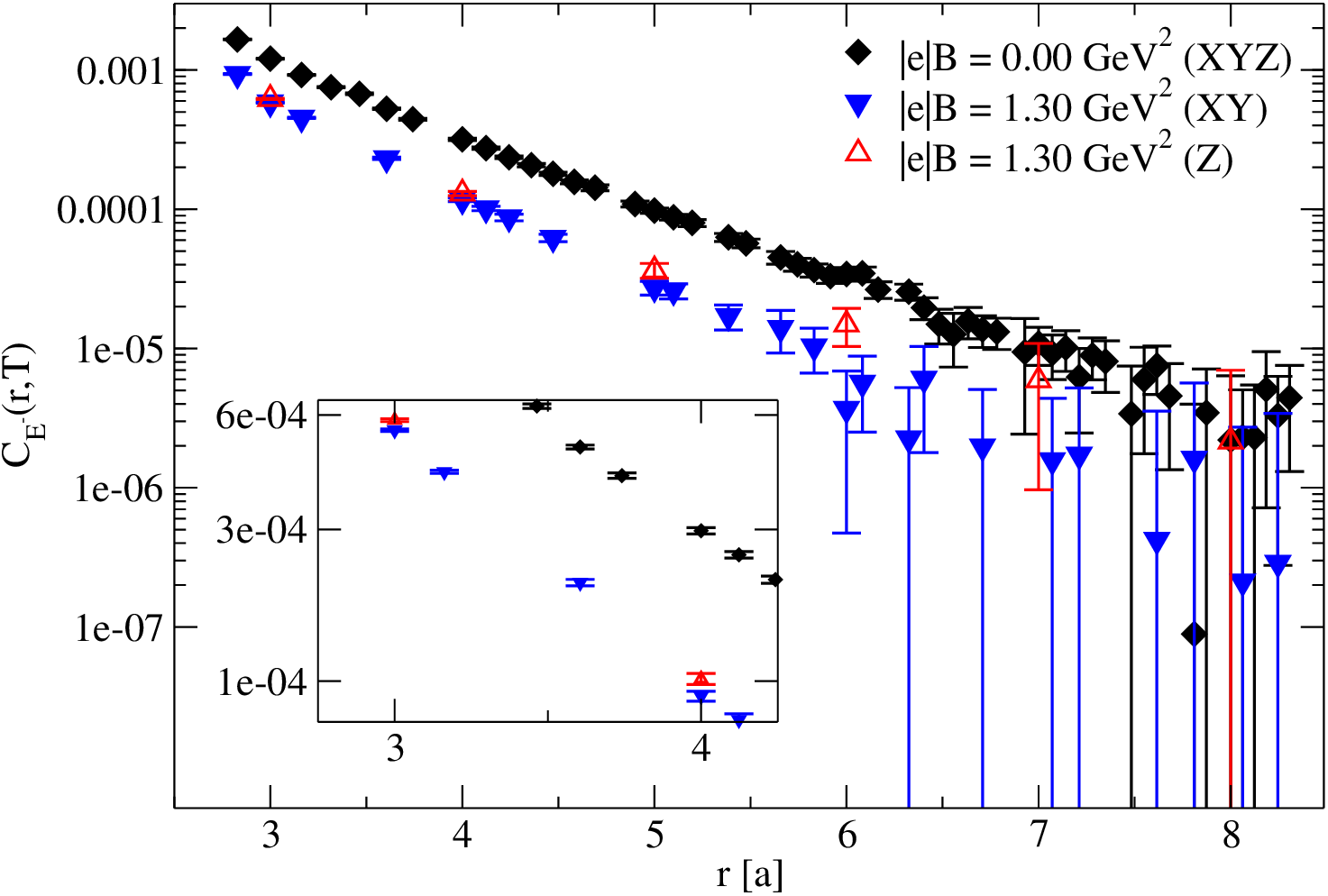}}
  \caption{Magnetic $C_{M^+}(r,T)$ (upper) and electric $C_{E^-}(r,T)$ (lower)
  correlators computed at $T\simeq 200~\mathrm{MeV}$ and with a magnetic field
  $|e|B\simeq 1.3~\mathrm{GeV}^2$. Correlators corresponding to separations
  parallel or orthogonal to the external magnetic field are denoted respectively
  by $Z$ and $XY$.  For comparison the results obtained for $e|B|=0$ are also
  displayed.}
  \label{fig:corrs}
\end{figure}

In Fig.~\ref{fig:corrs}, we show an example of the electric and magnetic
correlators for $T\simeq 200\,\mathrm{MeV}$ at $e|B|=0$ and $e|B|\simeq
1.3\,\mathrm{GeV}^2$. At zero magnetic field all possible orientations are
displayed, while for non-zero field only orientations perpendicular or parallel
to the magnetic field are considered. For this reason, the three curves in each
panel have different number of points.

The points at zero magnetic field approximately lay a single curve, indicating
that the lattice violations of rotational invariance are small. This holds true also
for correlators defined in the plane perpendicular to the magnetic field,
indicating that the residual $O(2)$ subgroup of $O(3)$ left unbroken by the
presence of $\mathbf{B}$ is well realized on the lattice for these correlator,
and in short, that lattice artefacts are small.

The external magnetic field is expected to modify the correlators both by
changing the screening masses and by inducing anisotropies in the correlators.
Fig.~\ref{fig:corrs} shows that both the electric and the magnetic correlators
approach zero faster when an external magnetic field is present, a fact that
implies that the screening masses are increasing functions of $B$. The
anisotropy of the correlators is in general not easy to observe in the high
temperature regime (see also \cite{nostro2}), since correlators decay very
quickly and it is possible to estimate them with enough relative accuracy only
for a short distance; nevertheless, for the relatively low $T$ and high $B$
case shown in Fig.~\ref{fig:corrs}, the anisotropy is present and more
pronounced in the magnetic correlator (whose signal is larger than the electric
one).  

\begin{figure}
  \subfigure{\includegraphics[width=\columnwidth, clip]{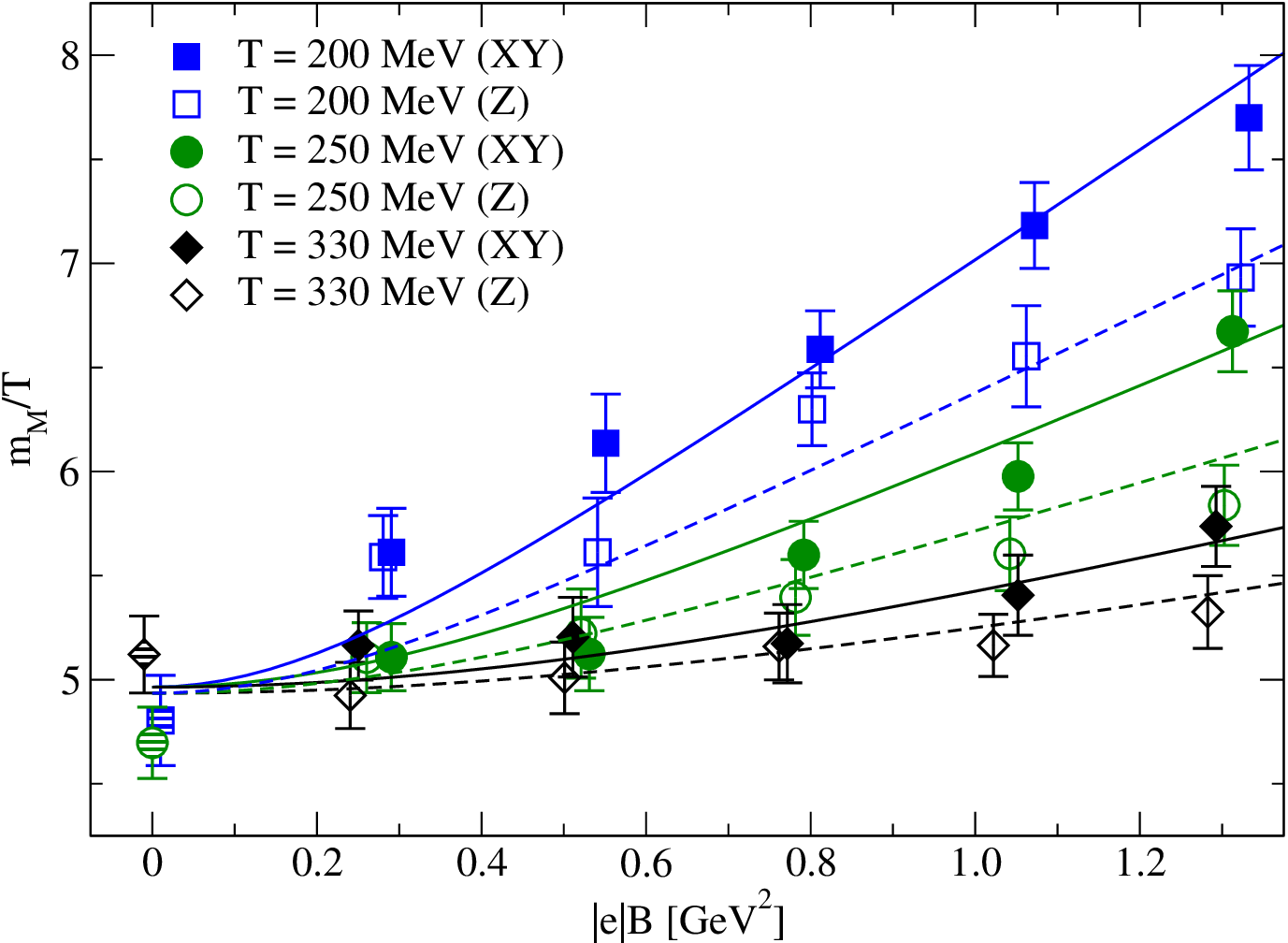}}
  \subfigure{\includegraphics[width=\columnwidth, clip]{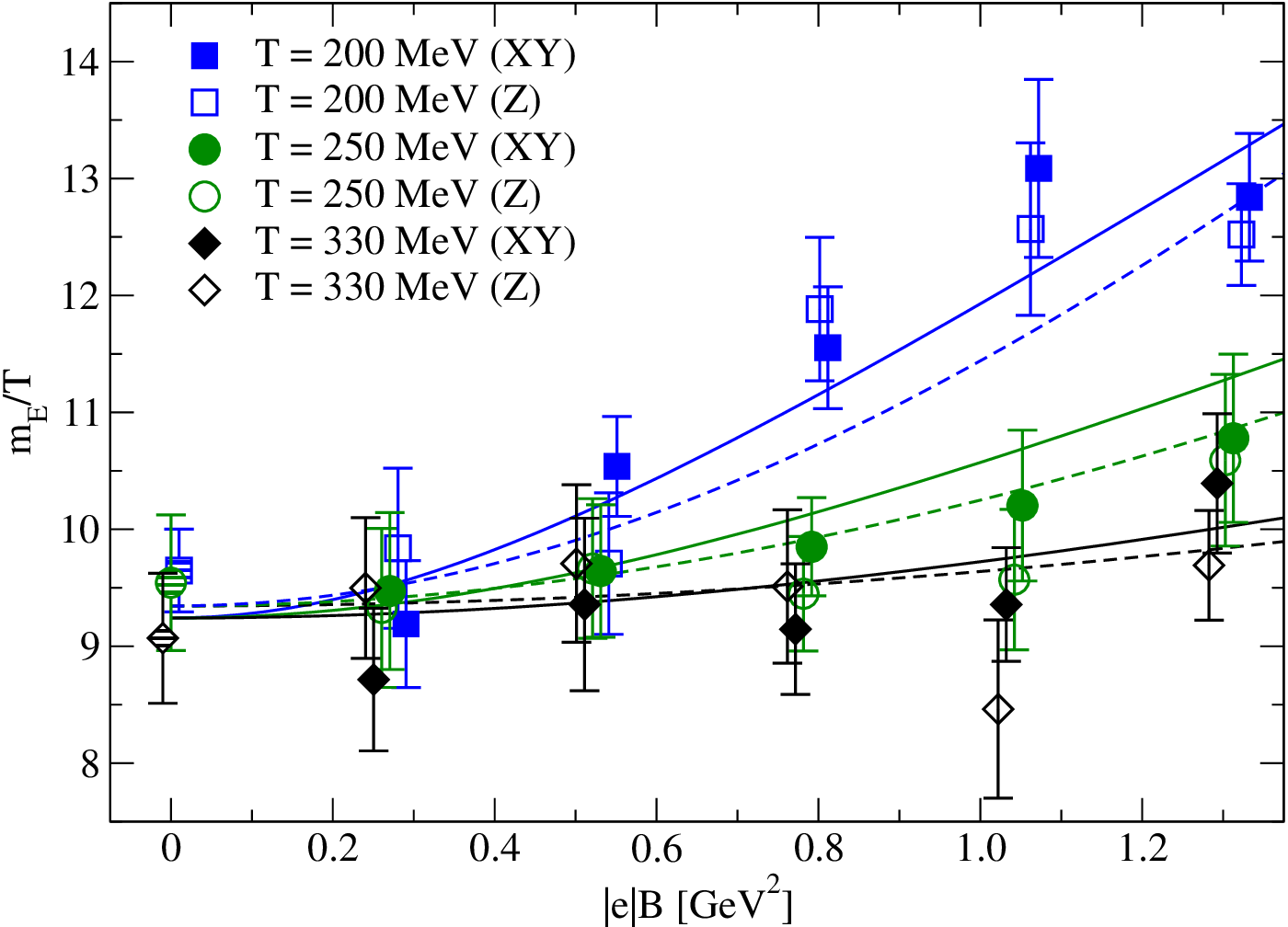}}
  \caption{Behavior of the ratios $m_{M}/T$ (upper) and $m_E/T$ (lower)
  as a function of the external field directed along $\hat{z}$.
  Data points  (slightly shifted on the horizontal axis to improve
  readability) are shown together with the best fit curves obtained
  by using the model in Eq.~\eqref{eq:model}.
  }
  \label{fig:m_vs_eB}
\end{figure}

\begin{figure}
  \subfigure{\includegraphics[width=\columnwidth, clip]{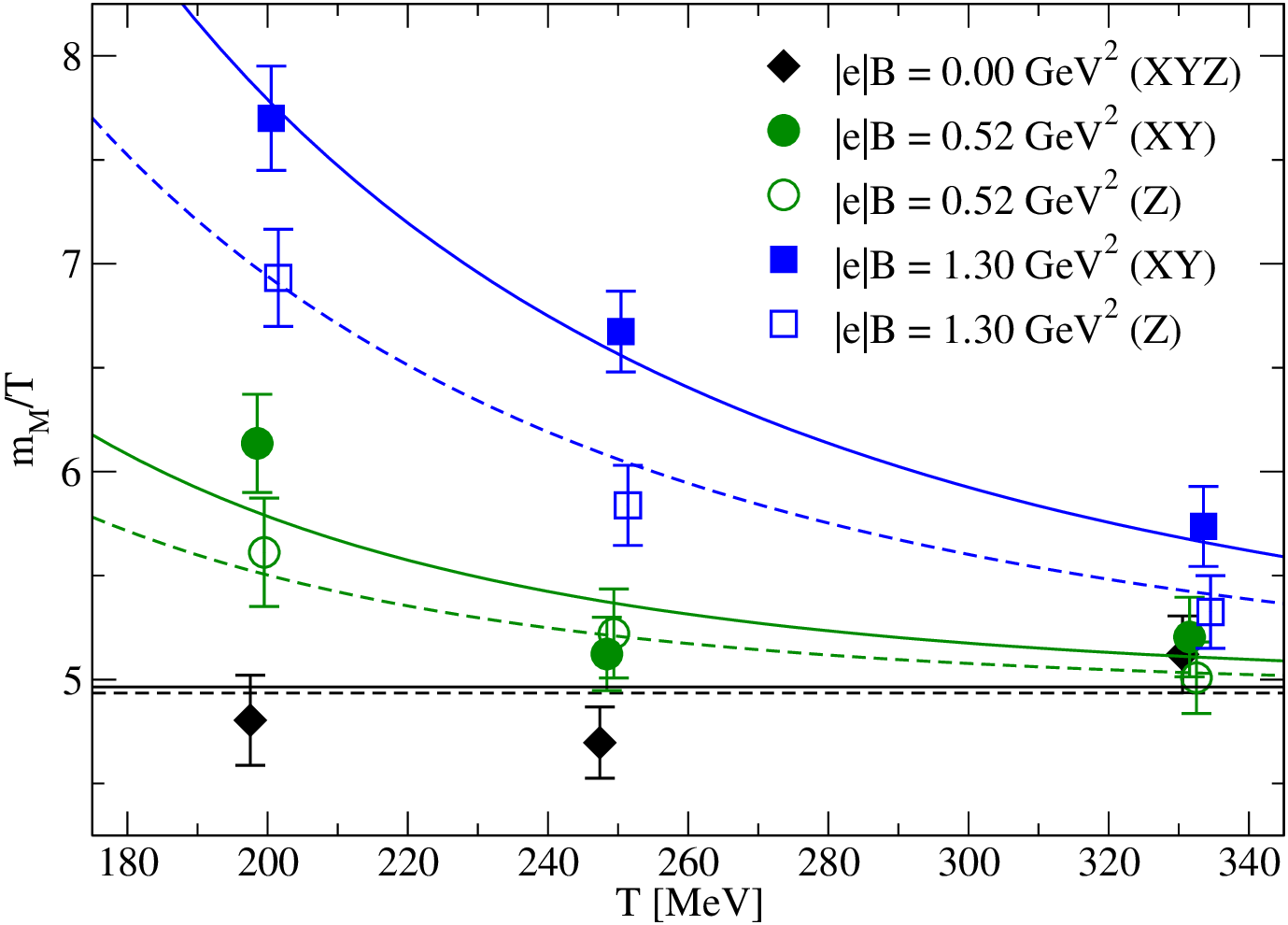}}
  \subfigure{\includegraphics[width=\columnwidth, clip]{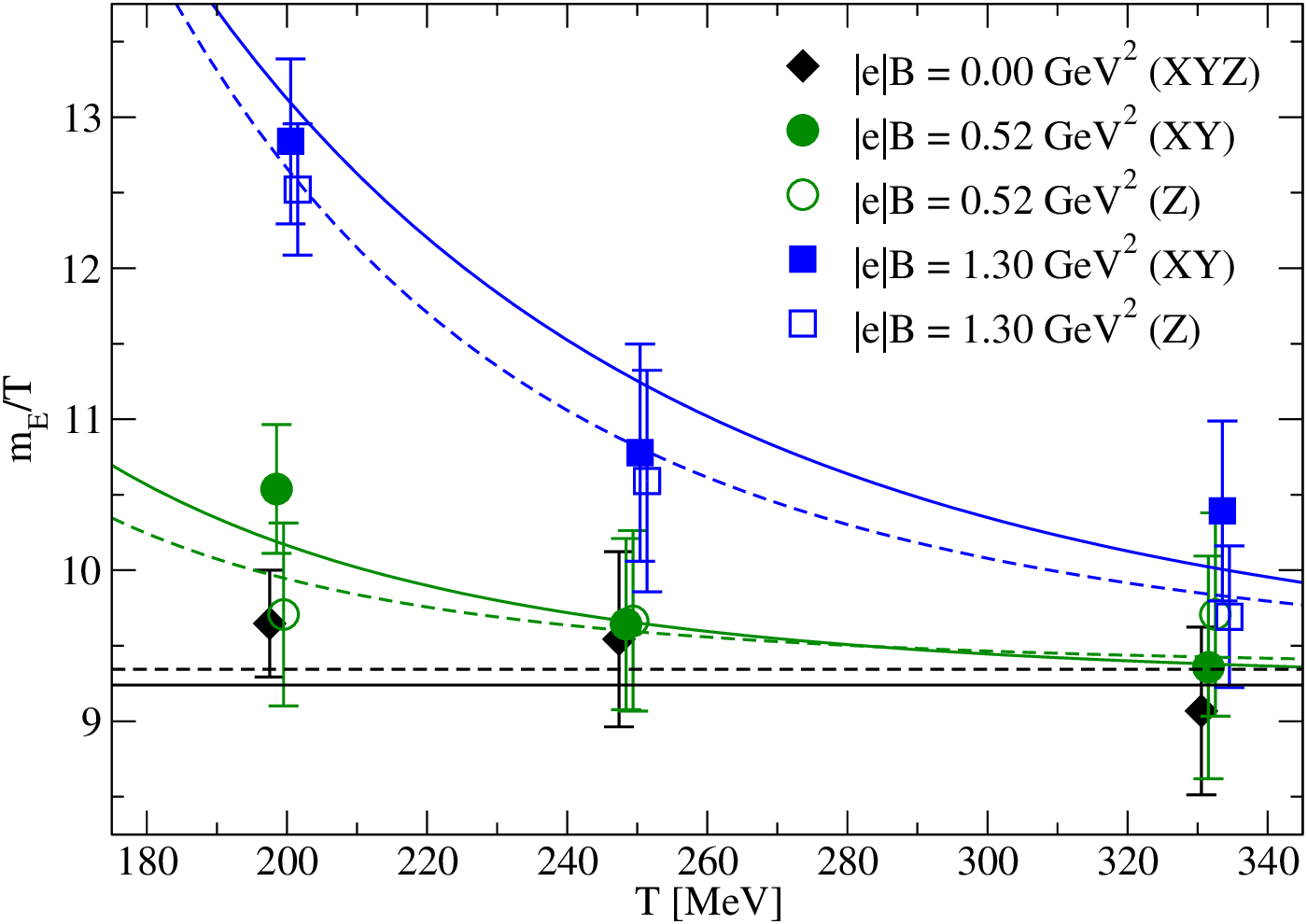}}
  \caption{Behavior of the ratios $m_{M}/T$ (upper) and $m_E/T$ (lower) as a
  function of the temperature.  Data points  (sligthly shifted on the horizontal
  axis to improve readability) are shown together with the best fit curves
  obtained by using the model in Eq.~\eqref{eq:model}.}
  \label{fig:m_vs_T}
\end{figure}

In order to determine the screening masses from the correlators, numerical data
have been fitted with the functional form in Eq.~\eqref{eq:mecorrsfit},
adopting a bootstrap approach to propagate the correlations between data.
Several fit intervals have been investigated in order to assess the stability
and reliability of the results and to estimate the systematic uncertainties
associated with the fitting procedure. In Table~\ref{tab:datapoints} we report the
numerical values obtained and in Figs.~\ref{fig:m_vs_eB}-\ref{fig:m_vs_T} we
show a graphical representation of their behavior as a function of $|e|B$ and
$T$.

At vanishing magnetic field we reproduced the known behavior of $m_E$ and
$m_M$: the electric screening mass is larger than the magnetic one and the
ratios $m_{M}/T$ and $m_E/T$ are remarkably insensitive to the value of the temperature,
something that \emph{a priori} would have been expected to hold only at much
higher temperatures.  Our results are in good agreement with the corresponding
temperatures and lattice spacing data presented in Ref.~\cite{Borsanyi:2015yka},
where the same discretization was used and to which we refer for an in depth
discussion of the $B=0$ case.

\begin{table}
  \begin{tabular}{c|c|c|c|c|c}
    \hline\hline
    \rule{0mm}{3.5mm}$T~\mathrm{[MeV]}$ & $|e|B~\mathrm{[GeV^2]}$ &  $m^{XY}_{M}/T$ & $m^{Z}_{M}/T$ & $m^{XY}_{E}/T$ & $m^{Z}_{E}/T$ \\
    \hline
    330 & 0.00 & 5.12(18) & 5.12(18) &  9.07(56) &  9.07(56) \\
    ''  & 0.26 & 5.16(17) & 4.92(16) &  8.71(61) &  9.50(60) \\
    ''  & 0.52 & 5.20(19) & 5.01(17) &  9.36(74) &  9.71(67) \\
    ''  & 0.78 & 5.17(19) & 5.16(16) &  9.15(56) &  9.51(66) \\
    ''  & 1.04 & 5.41(19) & 5.16(15) &  9.36(49) &  8.46(76) \\
    ''  & 1.30 & 5.77(19) & 5.32(17) & 10.39(60) &  9.69(47) \\
    \hline
    250 & 0.00 & 4.70(17) & 4.70(17) &  9.54(58) &  9.54(58) \\
    ''  & 0.26 & 5.11(16) & 5.11(17) &  9.47(67) &  9.33(68) \\
    ''  & 0.52 & 5.12(18) & 5.22(21) &  9.64(57) &  9.66(60) \\
    ''  & 0.78 & 5.60(16) & 5.39(18) &  9.85(42) &  9.45(49) \\
    ''  & 1.04 & 5.98(16) & 5.60(18) & 10.20(64) &  9.57(60)\\
    ''  & 1.30 & 6.67(19) & 5.84(19) & 10.78(72) & 10.59(73)\\
    \hline
    200 & 0.00 & 4.80(22) & 4.80(22) &  9.65(35) &  9.65(35) \\
    ''  & 0.26 & 5.61(21) & 5.59(20) &  9.19(54) &  9.84(68) \\
    ''  & 0.52 & 6.14(24) & 5.61(26) & 10.54(43) &  9.71(60) \\
    ''  & 0.78 & 6.59(18) & 6.30(17) & 11.55(52) & 11.88(61) \\
    ''  & 1.04 & 7.18(21) & 6.55(24) & 13.09(76) & 12.57(74) \\
    ''  & 1.30 & 7.70(25) & 6.93(23) & 12.84(55) & 12.52(43) \\
    \hline\hline
  \end{tabular}
  \caption{Screening masses obtained at three different temperatures and for
  several magnetic field intensities. Data at $|e|B=0$ have been obtained by
  averaging over all the spatial directions.}
  \label{tab:datapoints}
\end{table}

\begin{figure}[t]
  \includegraphics[width=\columnwidth,clip]{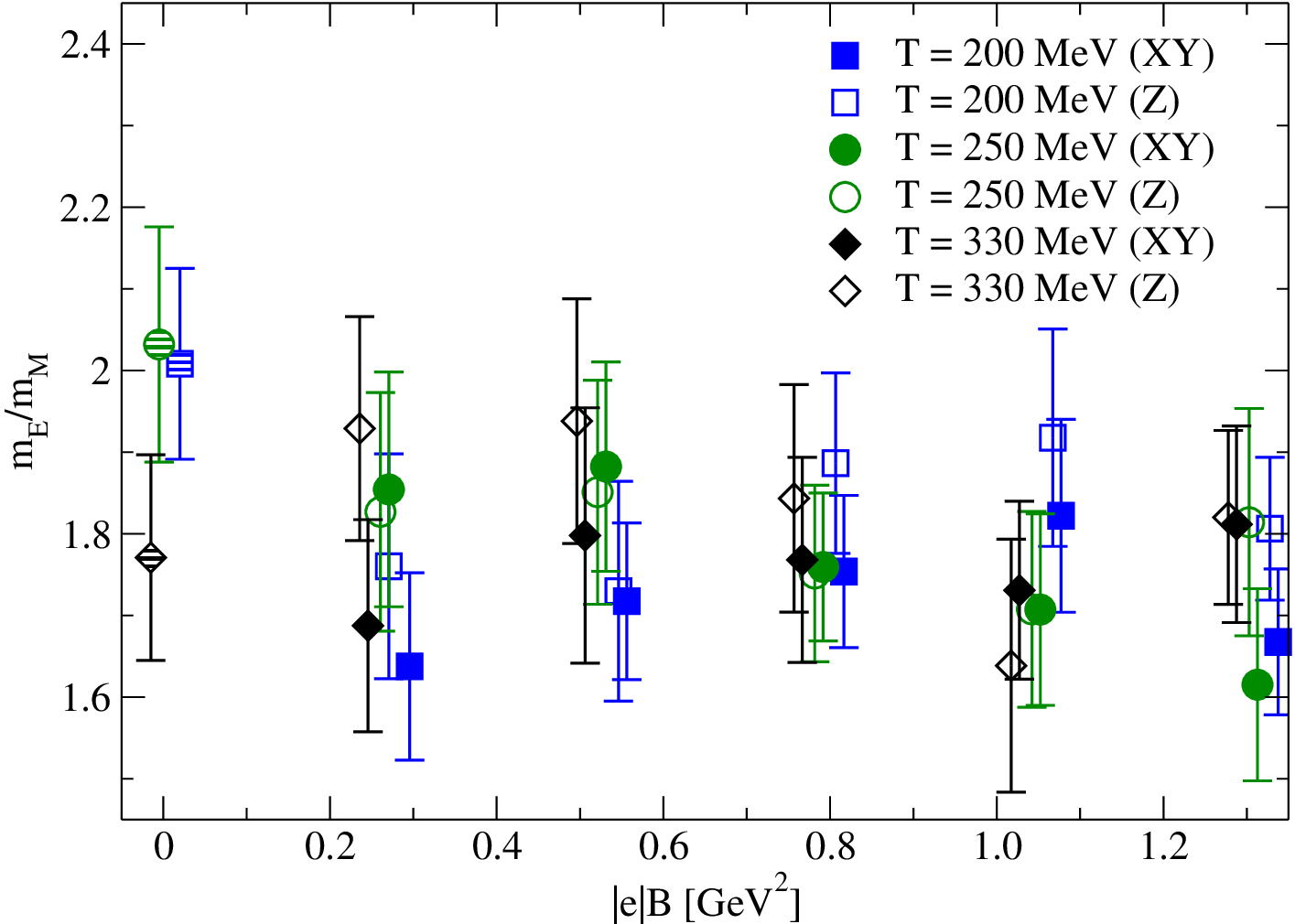}
  \caption{Determinations of the ratio $m_E/m_M$ for different
    temperatures and external field intensity.}
  \label{fig:massratio}
\end{figure}

As previously anticipated from Fig.~\ref{fig:corrs}, the effect of the magnetic
field is to increase both the magnetic and the electric screening mass, as
visible in Fig.~\ref{fig:m_vs_eB}. In both cases the growth is roughly
linear in the magnetic field and with similar slopes. Indeed the ratio
$m_E(T,B)/m_M(T,B)$ turns out to be independent of both the magnetic field
intensity and the temperature, as shown in Fig.~\ref{fig:massratio}. In
particular this means that that in the regime studied in this work the external
magnetic field does not change the usual $m_M<m_E$ hierarchy.

In the case of $m_M$ an anisotropy is observed, as could have been guessed by
Fig.~\ref{fig:corrs}, with the screening mass relative to the directions
orthogonal to the external field being larger than the one in the direction
parallel to the field.  In the case of the electric screening mass no such an
anisotropy is observed, however this could be due to the fact that data for
$m_E$ have larger relative error with respect to the ones for $m_B$. The
different accuracy of these estimates is a consequence of the relation
$m_E>m_M$: magnetic and electric correlators have similar statistical
(absolute) errors, but the extraction of $m_E$ is made difficult by the rapid
decrease of the electric correlator.

\begin{figure}
  \subfigure{\includegraphics[width=\columnwidth, clip]{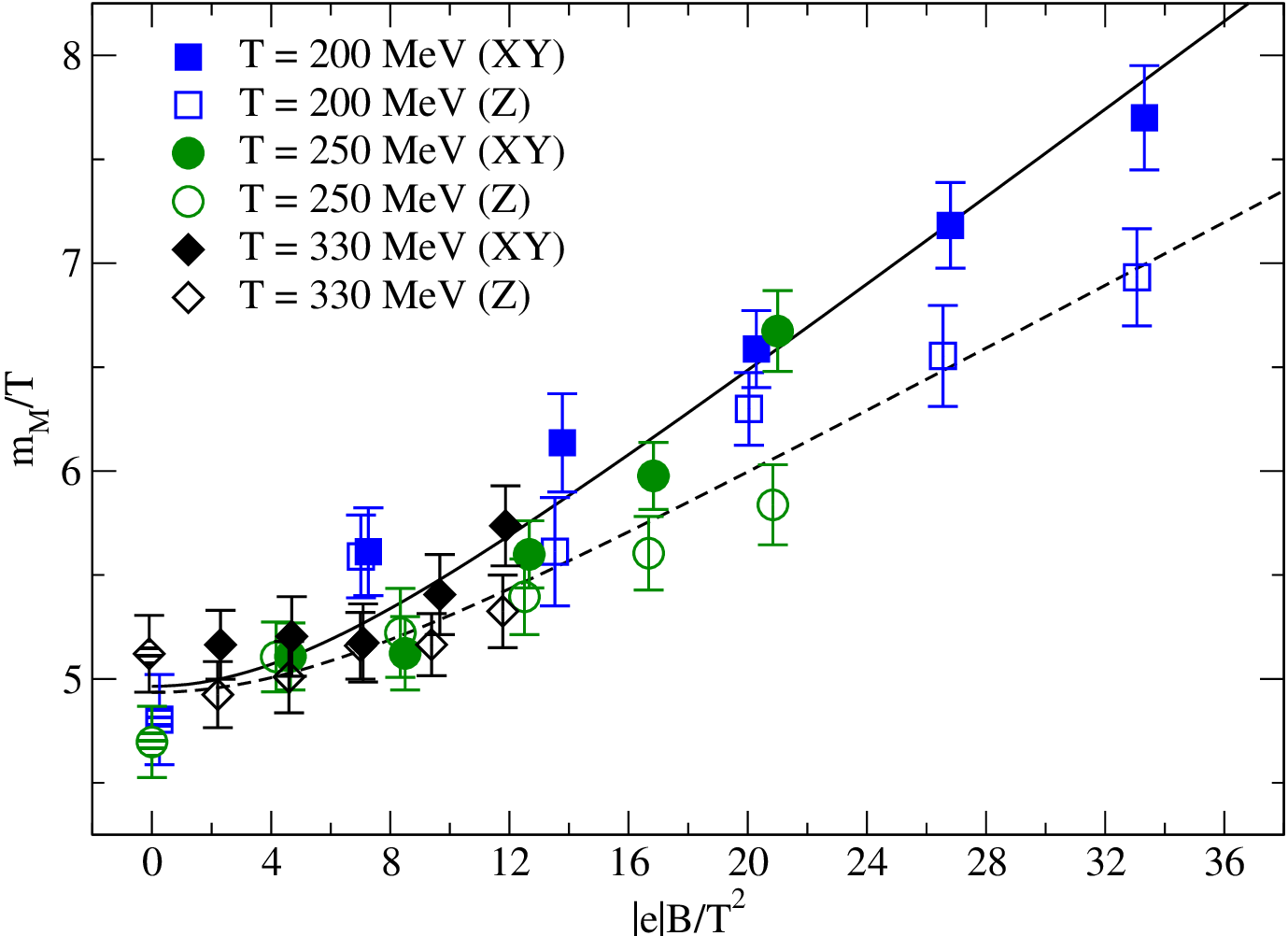}}
  \subfigure{\includegraphics[width=\columnwidth, clip]{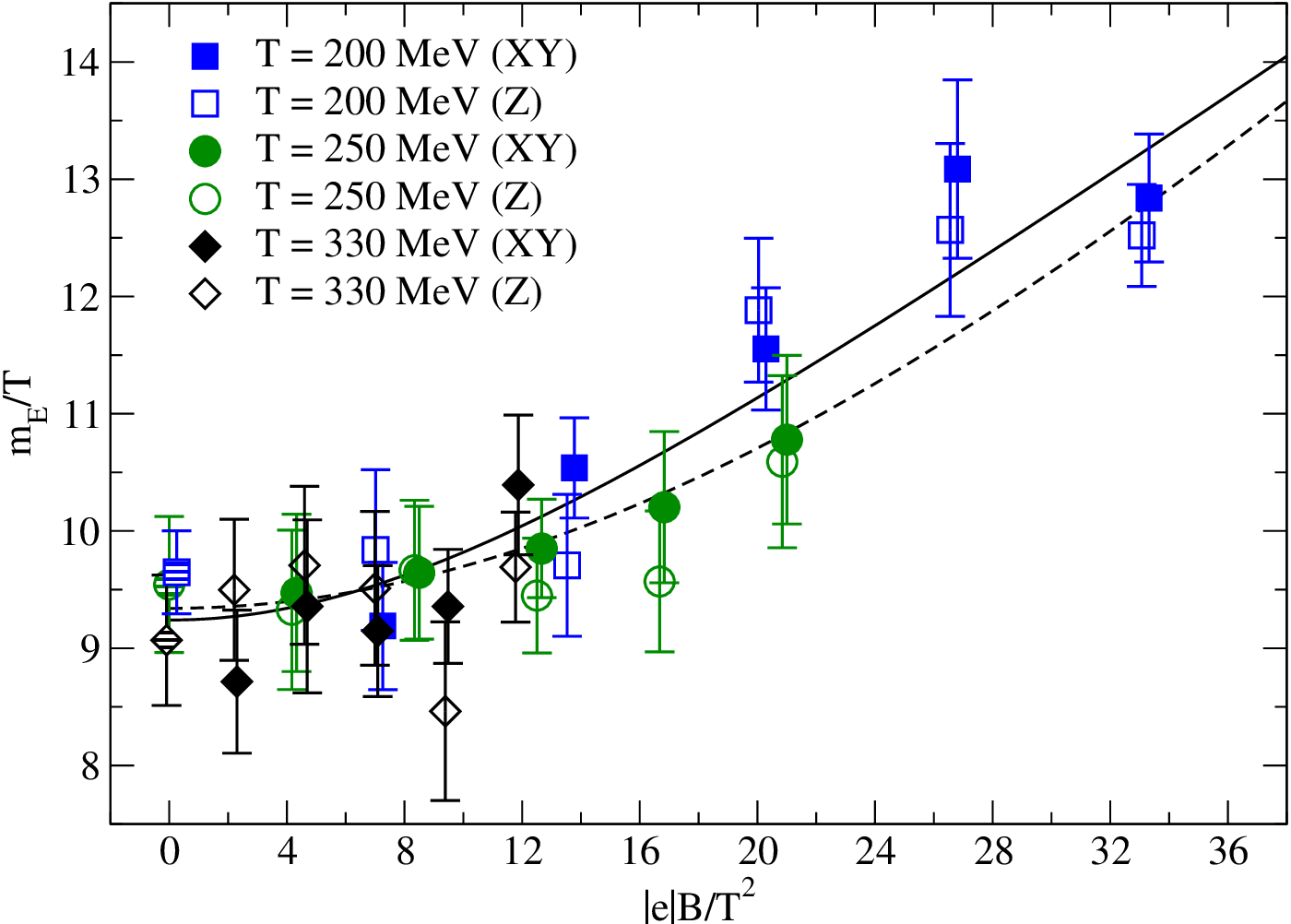}}
  \caption{Behavior of the ratios $m_{M}/T$ (upper) and $m_E/T$ (lower) as a
  function of $B/T^2$.  Data points are shown together with the best fit curves
  obtained by using the model in Eq.~\eqref{eq:model}.}
  \label{fig:m_vs_eB_T2}
\end{figure}

We now try to determine a functional form that well describes the $T$
and $B$ dependence of the screening masses.  From
Fig.~\ref{fig:m_vs_eB} and Fig.~\ref{fig:m_vs_T} it follows that the
main properties of the screening masses, at least in the explored
range of temperatures and magnetic field intensities, are:

\noindent
{\em i)} at $B=0$ the ratios $m_E/T$ and $m_M/T$ are independent of $T$; \\
{\em ii)}
for large magnetic field the screening masses grow linearly with $B$.

To these properties, it is reasonable to add the requirement that the
screening masses be analytic functions of the magnetic field
$\mathbf{B}$.  From that it follows that, in the limit of small
magnetic field intensity, the behaviors of $m_{E}$ and $m_{M}$ have to be
quadratic in $|e|B$.
In the high temperature phase it also seems reasonable to assume the only
relevant dimensional parameters to be the temperature and the magnetic field
intensity.  

Before going on we can explicitly check, in a model independent way,
that $B$ and $T$ are the only dimensional quantities that matter to
describe the behavior of the screening masses, by showing that the
ratios $m_E/T$ and $m_M/T$ depend only on the dimensionless
combination $B/T^2$. This is indeed the case, as can be appreciated from
Fig.~\ref{fig:m_vs_eB_T2}, where the $B^2$ behaviour for small
values of the magnetic field is also somehow more clearly visible than
in Figs~\ref{fig:m_vs_eB}-\ref{fig:m_vs_T}.

A simple ansatz that satisfies all the previous properties is
\begin{equation}\label{eq:model}
\frac{m_E^d}{T}=a^d_E\left[1+
c^d_{1;E}\frac{|e|B}{T^2}
\mathrm{atan}\left(\frac{c^d_{2;E}}{c^d_{1;E}}\frac{|e|B}{T^2}\right)\right]\ ,
\end{equation}
where $d$ denotes the spatial direction (i.e. $d=XY$ or $d=Z$) and an analogous
expression can be used for the magnetic screening mass.  This functional form
is analogous to the one used in Ref.~\cite{D'Elia:2011zu} for the case of the
dependence of the chiral condensate on the magnetic field.

The three parameters that enter the ansatz in Eq.~\eqref{eq:model} have simple
interpretations: $a_E$ is the $B=0$ value of the ratio $m_E/T$ (that is known
to be $T$-independent), $c_{1, E}$ is related to the slope of $m_E$ as a
function of $B$ at fixed temperature for large magnetic field intensities,
while $c_{2,E}$ is associated with the quadratic small-$B$ behavior of $m_E$.

The best fit values for the parameters entering Eq.~\eqref{eq:model}
for both $m_E$ and $m_M$ are reported in Table~\ref{tab:modelres}.
While for the magnetic screening masses reasonable values of the
$\chi^2$ test are obtained, for the case of the electric masses the
values of $\chi^2/\mathrm{d.o.f.}$ are somehow small, indicating that
we are using more parameters than needed to fit the data at the
current level of statistical accuracy. Indeed a simple linear
dependence on $|e|B$ is sufficient to describe the data for the
electric masses, however this is just a consequence of the large error
bars on the electric correlator.

\begin{table}
  \begin{tabular}{c|c|c|c|c}
    \hline\hline
    \rule{0mm}{3.5mm}  & $a$ &  $c_1$ &  $c_2$  & $\chi^2/\mathrm{d.o.f}$\\
    \hline
    \rule{0mm}{3.5mm}$m^{XY}_{M}$  & 4.964(82) & 0.137(19)$\times 10^{-1}$ & 0.141(55)$\times 10^{-2}$  &1.06 \\
    \rule{0mm}{3.5mm}$m^{Z~~}_{M}$ & 4.935(79) & 0.099(20)$\times 10^{-1}$ & 0.094(49)$\times 10^{-2}$  &1.10 \\
    \rule{0mm}{3.5mm}$m^{XY}_{E}$  & 9.24(21)  & 0.120(47)$\times 10^{-1}$ & 0.069(38)$\times 10^{-2}$  &0.63 \\
    \rule{0mm}{3.5mm}$m^{Z~~}_{E}$ & 9.34(20)  & 0.17(28) $\times 10^{-1}$ & 0.039(21)$\times 10^{-2}$  &0.85 \\
    \hline\hline
  \end{tabular}
  \caption{
  Best fit values for the parameters entering the functional form in
  Eq.~\eqref{eq:model}; in all cases $\mathrm{d.o.f.}=16$. }
  \label{tab:modelres}
\end{table}

\section{Conclusions}

In this study we have investigated the effects of a magnetic background on
color-screening phenomena taking place in the Quark-Gluon Plasma. To that
purpose, we have measured Polyakov loop correlators for various temperatures,
up to $T \simeq 330$ MeV, and uniform magnetic fields going up to
$|e|B\simeq 1.3$ GeV$^2$.  Our results have been obtained at a single value of the
lattice spacing, $a \sim 0.0989$~fm, and a refinement of the analysis, aimed
at a continuum limit extrapolation, should be performed in the future.

We have shown that the magnetic field induces an increase of both the magnetic
and the electric screening masses and, to some extent, also the appearance of
an anisotropy in Polyakov loop correlators. The masses increase linearly with
the magnetic field for moderate or large $B$ values (i.e. for $|e|B\gtrsim
0.2\,\mathrm{GeV}^2$) and a reasonable \emph{ansatz} can be given to describe
the connection of this regime with the expected quadratic behaviour for small
values of $B$, in which both screening masses are proportional to $T$ and
to a function of $B/T^2$. Indeed the influence of the magnetic field is
stronger at lower temperatures and asymptotically vanishes in the large $T$
limit.  On the other hand, the ratio of the magnetic to electric screening
masses turns to be independent of $B$, within errors, with the magnetic
screening mass always being smaller than the electric one.

The observed increase of the screening masses as a function of $B$ is in
qualitative agreement with perturbative computations~\cite{Alexandre:2000jc,
Bandyopadhyay:2016fyd,Bandyopadhyay:2017cle} and with the behavior
already observed below $T_c$: the magnetic background tends to suppress the
confining properties of the thermal medium below $T_c$, and to enhance the
screening of color interactions above $T_c$. In both cases, one can interpret
the effect in terms of the decrease of the 
pesudo-critical temperature $T_c$ as a function 
of $B$~\cite{Bali:2011qj}, so that
in the low temperature phase one approaches deconfinement as $B$ increases, 
while at high temperature one gets farther from the transition
and color screening gets stronger by increasing $B$. Following this
line of reasoning, no particular critical behavior is expected in the 
high temperature phase when approaching the 
large magnetic field limit, since the system will just become
more deconfined (i.e. color interactions will be more and more
screened); this is in contrast to what happens below $T_c$ or 
to what might happen even at $T = 0$~\cite{nostro2, Endrodi:2015oba}.  

The increasing of the screening masses induced by the presence of an
external magnetic field could in principle lead to a stronger
suppression of heavy quark bound states in peripheral heavy-ion
collisions and, more specifically, to a direct relation between
suppression and centrality. However, following the original argument
presented in Ref.~\cite{matsui}, an in deep discussion should consider
also the modifications of the radius of the heavy quarks bound states
as a function of $B$, since it is the reduction of the screening
length with respect to such radius which brings to the actual
suppression; a direct computation of quarkonia spectral functions in
the presence of magnetic background is surely something that should be
addressed by future lattice studies. Moreover, in order to assess the
relevance of our results to heavy ion phenomenology, one should first
of all know to which extent the magnetic field produced in non-central
heavy ion collisions survives the thermalization process, not to
mention all the other dynamical processes that make it difficult to
safely predict the fate of heavy quark bound states even in the
absence of external magnetic field.

\acknowledgments

We acknowledge PRACE for awarding us access to resource FERMI based in Italy at
CINECA, under project Pra09-2400 - SISMAF.

\end{document}